\documentclass[fleqn,usenatbib]{mnras}

\usepackage{newtxtext,newtxmath}

\usepackage[T1]{fontenc}

\DeclareRobustCommand{\VAN}[3]{#2}
\let\VANthebibliography\thebibliography
\def\thebibliography{\DeclareRobustCommand{\VAN}[3]{##3}\VANthebibliography}

\usepackage{graphicx}	
\usepackage{amsmath}	

\defcitealias{Tang2019}{T19}
\defcitealias{Tang2021a}{T21a}
\defcitealias{Tang2021b}{T21b}

\title[Blue extreme optical line emitting AGN]{Spectroscopy of an extreme [OIII] emitting active galactic nucleus at $\mathbf{z=3.212}$: implications for the reionisation era}

\author[M. Tang et al.]{
Mengtao Tang$^{1}$\thanks{E-mail: mengtao.tang@ucl.ac.uk}, 
Daniel P. Stark$^{2}$, 
Richard S. Ellis$^{1}$, 
St\'{e}phane Charlot$^{3}$, 
Anna Feltre$^{4}$, \newauthor
Alice E. Shapley$^{5}$ 
and Ryan Endsley$^{2}$
\\
\\
$^{1}$ Department of Physics and Astronomy, University College London, Gower Street, London WC1E 6BT, UK \\
$^{2}$ Steward Observatory, University of Arizona, 933 N Cherry Ave, Tucson, AZ 85721, USA \\
$^{3}$ Sorbonne Universit\'{e}, CNRS, UMR7095, Institut d'Astrophysique de Paris, F-75014, Paris, France \\
$^{4}$ INAF - Osservatorio di Astrofisica e Scienza dello Spazio di Bologna, Via P. Gobetti 93/3, 40129 Bologna, Italy \\
$^{5}$ Department of Physics and Astronomy, University of California, Los Angeles, 430 Portola Plaza, Los Angeles, CA 90095, USA 
}

\pubyear{2021}

\begin{document}
\label{firstpage}
\pagerange{\pageref{firstpage}--\pageref{lastpage}}
\maketitle

\begin{abstract}

Reionisation-era galaxies often display intense nebular emission lines, both in rest-frame optical ([O~{\small III}]+H$\beta$) and ultraviolet (UV; C~{\small III}], C~{\small IV}). How such strong nebular emission is powered remains unclear, with both active galactic nuclei (AGN) and hot stars considered equally viable. The UV continuum slopes of these early systems tend to be very blue ($\beta<-2$), reflecting minimal dust obscuration, young ages, and low metallicities. This contrasts with narrow-lined AGN at $z\sim2-3$, whose UV slopes are significantly redder ($\beta>-1$) than typical star-forming systems in the reionisation era. To investigate the properties of AGN in the reionisation era, we have conducted a search for potential examples of rare analogues with blue continua at intermediate redshift ($z\sim2-3$). Our goals are to determine whether AGN with intense line emission and blue continua exist and thereby to establish the range of rest-frame UV and optical line ratios in this population. In this paper we report the detection of a X-ray luminous AGN at $z=3.21$ (UDS-24561) with extreme [O~{\small III}]+H$\beta$ line emission (EW $=1300$~\AA) and a blue UV continuum slope ($\beta=-2.34$). MMT/Binospec and Keck/MOSFIRE spectra indicate rest-frame UV line ratios consistent with AGN photoionisation models and rest-frame optical lines with both a narrow component (FWHM $=154$~km$/$s) and extended broad wings (FWHM $=977$~km$/$s), consistent with outflowing gas. We describe how such objects can be identified in future {\it JWST} emission line surveys in the reionisation era, thereby providing a valuable census of AGN activity at $z>6$ and understanding their contribution to cosmic reionisation.

\end{abstract}





\section{Introduction} \label{sec:introduction}

Studying the reionisation of the intergalactic hydrogen is a major frontier in modern astrophysics, and provides important clues to understanding the history of cosmic structure formation \citep{Loeb2001}. The timeline of the reionisation process has been constrained by a series of observations undertaken over the last two decades. Measurement of the optical depth to Thomson scattering of the cosmic microwave background (CMB) radiation by the Planck team indicates a mid-point at $z=7.7$ \citep{PlanckCollaboration2020}. This is supported by a census of Ly$\alpha$ emitting galaxies at high redshift suggesting that the intergalactic medium (IGM) has a significant neutral hydrogen fraction ($x_{\rm{HI}}\sim50$~per~cent) at $z\sim7.5$ (see \citealt{Ouchi2020} for a review). Moreover, the redshift-dependent flux seen in the Ly$\alpha$ forest of quasar spectra implies a nearly fully ionised IGM and hence that the reionisation process is effectively complete by $z\simeq6$ \citep[e.g.,][]{Fan2006,McGreer2015}. 

However, the nature of the sources responsible for cosmic reionisation is still a matter of debate. Over the last decade, thousands of star-forming galaxies (SFGs) at $z>6$ have been detected in deep {\it Hubble Space Telescope} ({\it HST}) imaging surveys \citep[e.g.,][]{Bouwens2015a,Finkelstein2015,Oesch2018} and a popular view is that such sources provide the major contribution to reionisation \citep[e.g.,][]{Bouwens2015b,Robertson2015,Stanway2016,Dayal2018}. However, if all SFGs contribute equally, (i) the ultraviolet (UV) luminosity function of $z>6$ galaxies must extend to intrinsically feeble systems with $M_{\rm{UV}}\simeq-13$, and (ii) the fraction of ionising radiation that escapes into the IGM (the so-called `escape fraction', $f_{\rm{esc}}$) must be as high as $\simeq20$~per~cent (see \citealt{Stark2016} for a review). These requirements can be alleviated if rarer, more luminous systems make a more significant contribution, as suggested by the claimed rapid decline of the neutral fraction of the IGM over $6\lesssim z\lesssim7.5$ \citep[e.g.,][]{Naidu2020}. A possible explanation for this sudden end to reionisation is that many luminous sources may host active galactic nuclei (AGN) with both harder radiation fields \citep[e.g.,][]{Steidel2014,Stark2015,Feltre2016} and larger escape fractions ($f_{\rm{esc}}\sim75-100$~per~cent; e.g., \citealt{Cristiani2016,Grazian2018}). 

\begin{figure}
\begin{center}
\includegraphics[width=\linewidth]{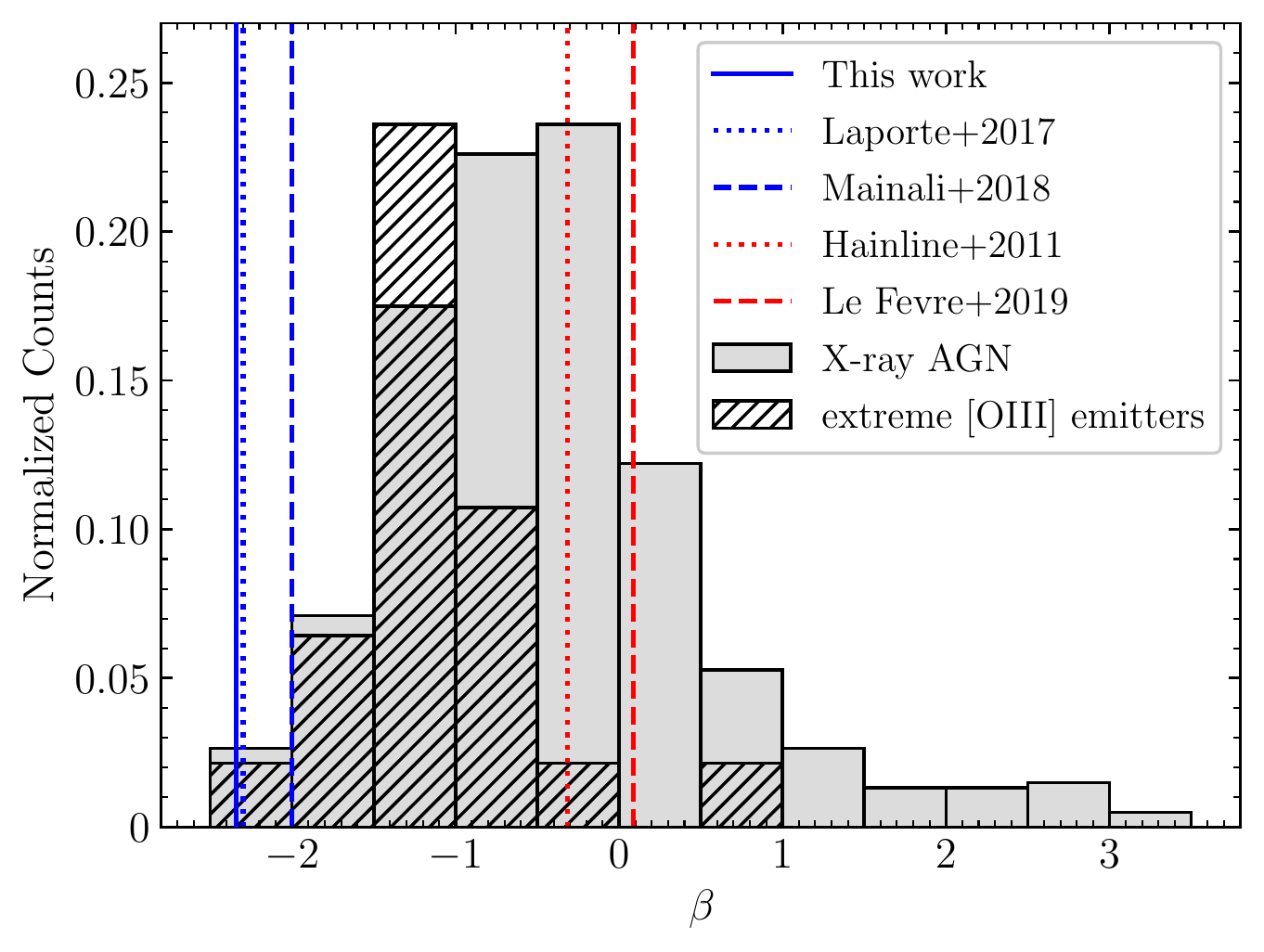}
\caption{UV slope ($\beta$) distribution of all the X-ray AGN at $z=1.3-3.7$ in the five CANDELS fields ($606$ objects; grey histogram) and that of the extreme [O~{\scriptsize III}] emitting AGN ($21$ objects; black hatched histogram), of which the y-axis is rescaled to normalised counts. UV slopes are derived from the \citet{Barro2019} catalogs by cross-matching with {\it Chandra} catalogs in CANDELS fields \citep{Xue2011,Xue2016,Nandra2015,Civano2016,Kocevski2018}. The UV slopes of X-ray AGN at $z=1.3-3.7$ are relatively red (median $\beta=-0.5$), consistent with the red UV slopes found in the stacked spectra of type II AGN at $z\sim2-4$ in \citet[][red dotted line]{Hainline2011} and \citet[][red dashed line]{LeFevre2019}, while the extreme [O~{\scriptsize III}] emitting AGN show bluer UV slopes (median $\beta=-1.0$). On the other hand, the N~{\scriptsize V} emitting AGN at $z>7$ (\citealt{Laporte2017}, blue dotted line; \citealt{Mainali2018}, blue dashed line) and the $z=3.21$ AGN studied in this paper, UDS-24561 (blue solid line), have much bluer UV slopes of $\beta<-2.0$.}
\label{fig:beta_dist}
\end{center}
\end{figure}

At present, there is only indirect evidence supporting the hypothesis of AGN activity in galaxies at $z>6$. High ionisation UV emission lines (e.g., N~{\small V}~$\lambda\lambda1238,1243$, C~{\small IV}~$\lambda\lambda1548,1550$) have been detected in a handful of star-forming galaxies \citep[e.g.,][]{Tilvi2016,Hu2017,Laporte2017,Mainali2018,Sobral2019,Endsley2021b,Onoue2021}. Although many AGN have been found in SFGs at $z\sim2-4$ \citep[e.g.,][]{Hainline2011}, it seems the limited number of corresponding reionisation-era systems with potential AGN signatures show very different properties. This may make it questionable to identify AGN in $z>6$ systems using the predominant spectral features seen in $z\sim2-4$ AGN. Specifically, SFGs in the reionisation era often display rest-frame optical emission lines with extremely high equivalent widths (EWs), with an average [O~{\small III}]+H$\beta$ EW $\simeq700$~\AA\ at $z\sim7-8$ \citep{Labbe2013,DeBarros2019,Endsley2021a}. Such intense lines are seldom seen in lower redshift SFGs with AGN \citep[e.g.,][]{Tang2019}. Likewise, $z>6$ galaxies have very steep UV continuum slopes ($\beta<-2.0$; e.g., \citealt{Bouwens2012,Bouwens2014,Finkelstein2012}) indicating metal-poor stellar populations with little dust. On the other hand, AGN at $z\sim2-4$ present much redder UV slopes ($\beta>-1$, Fig.~\ref{fig:beta_dist}; e.g., \citealt{Hainline2011,LeFevre2019}), presumably because they are hosted by more massive galaxies with increased dust extinction \citep[e.g.,][]{Hainline2012}. A key question is whether these differences suggest that the nature and properties of AGN evolve significantly between $z\sim2-3$ and $z>6$, and that a population of blue, extreme optical line emitting galaxies harbouring AGN are somehow more abundant in the reionisation era. 

The above question has motivated us to search for examples of AGN activity in $z\simeq1-3$ SFGs with blue UV slopes and extreme optical line emission. To accomplish this, we exploit our large spectroscopic survey of extreme [O~{\small III}] emitting galaxies at $z=1.3-3.7$ at rest-frame optical \citep[][hereafter \citetalias{Tang2019}]{Tang2019} and UV wavelengths \citep[][hereafter \citetalias{Tang2021a,Tang2021b}]{Tang2021a,Tang2021b}, targeting selected systems for evidence of AGN activity using diagnostic UV and optical emission lines in the context of photoionisation models \citep[e.g.,][]{Allen2008,Feltre2016,Jaskot2016,Nakajima2018,Hirschmann2019,Plat2019}. This enables us to distinguish between spectra powered by AGN and massive stars. In this paper, we present the case of a $z=3.212$ AGN with extremely large EW [O~{\small III}] emission with a steep UV continuum slope whose properties are thus similar to SFGs in the reionisation era \citep[e.g.,][]{Endsley2021a}. Our ultimate goal is to build a statistical sample of such $z\simeq1-3$ AGN which may guide future spectroscopic searches for AGN activity in the reionisation era using future facilities such as the {\it James Webb Space Telescope} ({\it JWST}) or new generation of $30$ metre class telescopes which can target the relevant UV and optical nebular emission lines.

The organisation of this paper is as follows. We describe our target selection and the spectroscopic observations in Section~\ref{sec:observations}. The emission line measurements and the associated line ratio diagnostics that support evidence of AGN activity are discussed in Section~\ref{sec:results}. Finally, we discuss the implications of our results for the contribution of AGN to the reionisation era in Section~\ref{sec:discussion}. We adopt a $\Lambda$-dominated, flat universe with $\Omega_{\Lambda}=0.7$, $\Omega_{\rm{M}}=0.3$, and $H_0=70$~km~s$^{-1}$~Mpc$^{-1}$. All magnitudes in this paper are quoted in the AB system \citep{Oke1983}, and all EWs are quoted in the rest frame.


\section{Observations} \label{sec:observations}

In this work, we aim to locate and study the rest-frame UV and optical spectra of AGN with extremely large EW [O~{\small III}] line emission and blue UV slopes. Our candidates will be chosen from a parent sample of extreme [O~{\small III}] emitting AGN at $z=1.3-3.7$. We describe the pre-selection of targets in Section~\ref{sec:selection}, and the spectroscopic observations in Section~\ref{sec:binospec}, \ref{sec:mosfire}, and \ref{sec:hst_grism}. 

\subsection{Pre-selection of extreme line emitting AGN with blue UV continuum slopes at $\mathbf{z=1.3-3.7}$} \label{sec:selection}

The first step of this study is to locate AGN candidates with extreme EW [O~{\small III}] line emission and blue UV slopes for spectroscopic follow-up. In \citetalias{Tang2019}, we discuss the selection of a sample of extreme [O~{\small III}] emitting objects at $z=1.3-3.7$ in the Cosmic Assembly Near-infrared Deep Extragalactic Legacy Survey (CANDELS) fields \citep{Grogin2011,Koekemoer2011} and their near-infrared (rest-frame optical) spectra. We direct the reader to \citetalias{Tang2019} for details. In brief, extreme [O~{\small III}] emitters were selected to have large rest-frame [O~{\small III}]~$\lambda\lambda4959,5007$ EWs with values $\simeq300-3000$~\AA\ chosen to match the range common in reionisation-era sources \citep[e.g.,][]{Endsley2021a}. The [O~{\small III}] EWs were inferred from {\it HST} grism spectra (at $z=1.3-2.4$; \citetalias{Tang2019}) or the $K$-band flux excess (at $z=3.1-3.7$; Tang et al. in preparation) using the 3D-HST catalogs \citep{Brammer2012,Skelton2014,Momcheva2016}. 

To identify which are AGN we cross-match our $z=1.3-3.7$ extreme [O~{\small III}] emitter sample to deep {\it Chandra} X-ray source catalogues in the All-Wavelength Extended Groth Strip International Survey (AEGIS; \citealt{Nandra2015}), the Cosmic Evolution Survey (COSMOS; \citealt{Civano2016}), the Great Observatories Origins Deep Survey North (GOODS-N; \citealt{Xue2016}) and South (GOODS-S; \citealt{Xue2011}), and the Ultra Deep Survey (UDS; \citealt{Kocevski2018}) fields. The {\it Chandra} X-ray data used in the five CANDELS fields were reduced following the prescription described in \citet{Laird2009} and \citet{Nandra2015}. The depths of the X-ray data are $800$ ks in AEGIS, $160$ ks in COSMOS, $2$ Ms in GOODS-N, $4$ Ms in GOODS-S, and $600$ ks in UDS, resulting in hard band ($2-10$ keV) flux limits (over $>1$~per~cent of the survey area) of $2.5\times10^{-16}$, $1.8\times10^{-15}$, $5.9\times10^{-17}$, $5.5\times10^{-17}$, and $6.5\times10^{-16}$~erg~s$^{-1}$~cm$^{-2}$, respectively. We match the coordinates of the extreme [O~{\small III}] emitting objects (from the 3D-HST catalogs) to the X-ray source catalogs using a $1.0$~arcsec search radius. There are $21$ sources in our sample found to have X-ray counterparts within $1.0$~arcsec. 

We next examine whether the X-ray luminosities ($L_{\rm{X}}$) of the $21$ sources are consistent with those of X-ray AGN, which typically have $L_{\rm{X}}\gtrsim10^{42}$~erg~s$^{-1}$ at rest-frame $2-10$~keV. The X-ray luminosities of our sources were estimated by fitting the X-ray spectra assuming a power law with index $\Gamma=1.7-1.8$ for intrinsic AGN spectra (AEGIS: \citealt{Buchner2015}; COSMOS: \citealt{Marchesi2016}; GOODS-N: \citealt{Xue2016}; GOODS-S: \citealt{Xue2011}; UDS: \citealt{Kocevski2018}). The absorption corrected rest-frame $2-10$~keV luminosities ($L_{2-10\ \rm{keV}}$) of the $21$ X-ray extreme [O~{\small III}] emitting sources in our sample range from $1.1\times10^{42}$ to $1.3\times10^{45}$~erg~s$^{-1}$. Since all the $21$ sources have $L_{2-10\ \rm{keV}}>10^{42}$~erg~s$^{-1}$, each is consistent with being an AGN.

The UV slope distribution of the $21$ extreme [O~{\small III}] emitting AGN at $z=1.3-3.7$ is compared in Fig.~\ref{fig:beta_dist} (black hatched histogram) to that of the total X-ray AGN sample in the CANDELS fields (grey histogram). Using the parent $z=1.3-3.7$ extreme [O~{\small III}] emitter sample in \citetalias{Tang2019}, we estimate the fraction of extreme [O~{\small III}] emitting X-ray AGN as a function of UV slope. About $0.3$, $0.5$, $15$, and $26$~per~cent of the extreme [O~{\small III}] emitters with UV slopes $\beta<-2.0$, $=-2.0$ to $-1.5$, $=-1.5$ to $-1.0$, and $>-1.0$ harbour X-ray AGN, respectively. We note that we do not account for systems such as faint X-ray AGN or infrared AGN which the X-ray luminosities may be below the detection limit of current X-ray surveys, so the AGN fractions could be larger. Although extreme [O~{\small III}] emitting AGN have relatively red (median $\beta=-1.0$) UV slopes, they are on average bluer than those of the total X-ray AGN sample (median $\beta=-0.5$). As the type II AGN continuum should not contribute significantly at rest-frame UV wavelengths \citep{Assef2010}, it appears that the host galaxies of extreme [O~{\small III}] emitters are less dusty or more metal-poor than those of the majority of type II AGN at $z\sim1-3$.

Our goal in this paper is to identify AGN at intermediate redshifts that are similar to the $z\simeq7-9$ systems thought to harbour AGN in \citet{Laporte2017} and \citet{Mainali2018}. Those objects are characterised by even bluer UV continuum slopes ($\beta<-2.0$; Fig.~\ref{fig:beta_dist}) and very large [O~{\small III}]+H$\beta$ EWs ($>900$~\AA\ in rest-frame) inferred from their extremely red {\it Spitzer}/IRAC $[3.6]-[4.5]$ colours. Among the $21$ sources in our sample, only one object, UDS-24561 (R.A. $=02$:$17$:$44.462$, Decl. $=-05$:$11$:$38.25$; Fig.~\ref{fig:image}), satisfies the joint criteria of $\beta<-2.0$ and rest-frame EW$_{\rm{[OIII]+H}\beta}>900$~\AA. By measuring the emission lines shown in the low resolution ($R=130$) {\it HST} grism spectrum (\citealt{Momcheva2016}; see Section~\ref{sec:line}), the estimated spectroscopic redshift of UDS-24561 is $z=3.21$. We now examine this source in greater detail.

The multi-wavelength spectral energy distribution (SED) of UDS-24561 is presented in Fig.~\ref{fig:sed}, which is extracted from the \citet{Skelton2014} photometry catalogs. This object is extremely luminous with absolute UV magnitude $M_{\rm{UV}}=-22.4$, i.e., about five times brighter than the characteristic $M^*_{\rm{UV}}$ at $z\sim3$ ($M^*_{\rm{UV}}\simeq-20.6$; see \citealt{Parsa2016}, and references therein). The UV continuum slope of UDS-24561 is $\beta=-2.34$ and was computed by fitting a power law ($f_{\lambda}\propto\lambda^{\beta}$) to the broadband fluxes at rest-frame $1250-2600$~\AA\ \citep{Calzetti1994}. The intense $K$-band flux excess indicates a large rest-frame [O~{\small III}]+H$\beta$ EW ($\simeq1300$~\AA; see Section~\ref{sec:line} for more accurate measurements). Both properties are comparable to those of the putative AGN at $z\simeq7-9$ discussed by \citet{Laporte2017} and \citet{Mainali2018}. The absorption corrected $L_{2-10\ \rm{keV}}$ of this object in the \citet{Kocevski2018} catalog is $2.75\times10^{44}$~erg~s$^{-1}$, indicating that it is extremely luminous. The stellar mass inferred from BEAGLE \citep{Chevallard2016} SED fitting is $M_{\star}=2.0\times10^{10}\ M_{\odot}$. We note that the current version of the BEAGLE code does not permit including an AGN conponent, and since we primarily focus on identifying blue, extreme line emitting systems harbouring AGN in this paper, we do not discuss the SED fitting parameters in detail. We obtained the optical and near-infrared (rest-frame UV and optical) spectra of UDS-24561 via ground-based observations and {\it HST} archival data. We discuss the spectroscopic observations in the following subsections. 

\subsection{MMT/Binospec spectroscopy} \label{sec:binospec}

The optical spectrum of UDS-24561 was obtained using the multi-slit spectrograph Binospec \citep{Fabricant2019} on the MMT in order to measure the rest-frame UV emission lines. We designed one mask in the CANDELS/UDS field (centered at R.A. $=02$:$17$:$26.8$ and Decl. $=-05$:$17$:$36$, with position angle PA $=-81^{\circ}$) primarily targeting on UDS-24561. We filled the mask with non-AGN extreme [O~{\small III}] emitting galaxies at $z=1.3-3.7$, thereby continuing an ongoing survey targeting rest-frame UV metal emission lines \citepalias{Tang2021a} and the Ly$\alpha$ emission line \citepalias{Tang2021b} in these systems. We observed this mask between 2020 October and November with a total on-target integration time of $28800$ seconds during average seeing of $0.9$~arcsec. Optical spectra of the targets were taken using the $270$~lines~mm$^{-1}$ grism blazed at $5.5^{\circ}$, with wavelength coverage from $3850$ to $9000$~\AA. The slit width was set to $1.0$~arcsec, which results in a spectral resolution of $R=1340$. This mask contains two slit stars to compute the absolute flux calibration, and we also observed spectrophotometric standard stars at a similar airmass in order to correct the instrument response. 

The Binospec spectra were reduced using the publicly available data reduction pipeline\footnote{\url{https://bitbucket.org/chil_sai/binospec}} \citep{Kansky2019}. The pipeline performs flat-fielding, wavelength calibration, sky subtraction, and then the 2D spectra extraction. We created 1D spectra from the reduced 2D spectra using a boxcar extraction, with the extraction aperture matched to the spatial profile of the object. The atmospheric extinction and instrumental response were corrected using the sensitivity curve derived from observations of standard stars. We performed slit loss correction following the similar procedures in \citetalias{Tang2019}. We derived the spatial profile of each target from its {\it HST} F814W postage stamp, and computed the fraction of the light within the slit to that of the total spatial profile. The flux of each spectrum was then divided by the in-slit light fraction measured for each object. Finally, the absolute flux calibration was performed using observations of slit stars, by comparing the slit-loss corrected count rates of slit star spectra with the flux in the \citet{Skelton2014} catalogues.

\subsection{Keck/MOSFIRE spectroscopy} \label{sec:mosfire}

We also obtained a near-infrared spectrum for UDS-24561 using the Multi-object Spectrometer for Infrared Exploration (MOSFIRE; \citealt{McLean2012}) on the Keck I telescope. We designed one multi-slit mask (centered at R.A. $=02$:$17$:$41.25$ and Decl. $=-05$:$12$:$52.26$, with position angle PA $=71^{\circ}$) which primarily targets on UDS-24561. The mask was filled with non-AGN extreme [O~{\small III}] emitting galaxies at $z=2.1-3.7$ following the survey strategy discussed in \citetalias{Tang2019}. We observed this mask on 2021 March 3 and 4, with a total on-target integration time of $2880$ seconds and an average seeing of $0.8$~arcsec. The spectra were taken in the $K$-band with wavelength coverage from $1.954$ to $2.397\ \mu$m, aiming to measure the H$\beta$ and [O~{\small III}] emission lines for UDS-24561, as well as H$\alpha$ or H$\beta$ and [O~{\small III}] emission lines for the fillers. The slit width was set to $0.7$~arcsec, resulting in a spectral resolution of $R=3610$. We also placed two slit stars on the mask for absolute flux calibration, and observed A0V stars to derive the response spectra and correct for telluric absorption. 

The MOSFIRE spectra were reduced using the publicly available {\footnotesize PYTHON}-based data reduction  pipeline\footnote{\url{https://keck-datareductionpipelines.github.io/MosfireDRP}} (DRP). The pipeline performs flat-fielding, wavelength calibration, and background subtraction before 2D spectra extraction. We created 1D spectra from the reduced 2D spectra using a boxcar extraction. The telluric absorption and instrumental response were corrected using observations of A0V stars. Slit loss correction of each target was performed using the in-slit light fraction computed from its {\it HST} F160W image. We then performed the absolute flux calibration using observations of slit stars, by comparing the observed fluxes of star spectra with the broadband photometry.

\subsection{\textit{HST} WFC3/G141 spectroscopy} \label{sec:hst_grism}

In addition to the MMT/Binospec and the Keck/MOSFIRE $K$-band observations, we use the emission line measurements in the {\it HST} Wide Field Camera 3 (WFC3) G141 slitless grism spectrum of UDS-24561 from the 3D-HST survey, which is described in detail in \citet{Momcheva2016}. The {\it HST} WFC3/G141 grism has spectral coverage from $1.1$ to $1.7\ \mu$m (corresponding to ground-based $J$ and $H$ bands) and a spectral resolution of $R=130$. The grism images were first reduced by removing satellites trails and earthshine, followed by flat-fielding and background subtraction. Then the exposures were combined by interlacing into output mosaic images used for the spectral extractions. After that a contamination model was created to account for overlapping spectra due to neighboring sources, and the 2D spectra of individual objects were extracted from the interlaced mosaic images and contamination were subtracted. Finally, the 1D spectra were extracted using the optimal extraction procedures \citep{Horne1986}.


\begin{figure}
\begin{center}
\includegraphics[width=\linewidth]{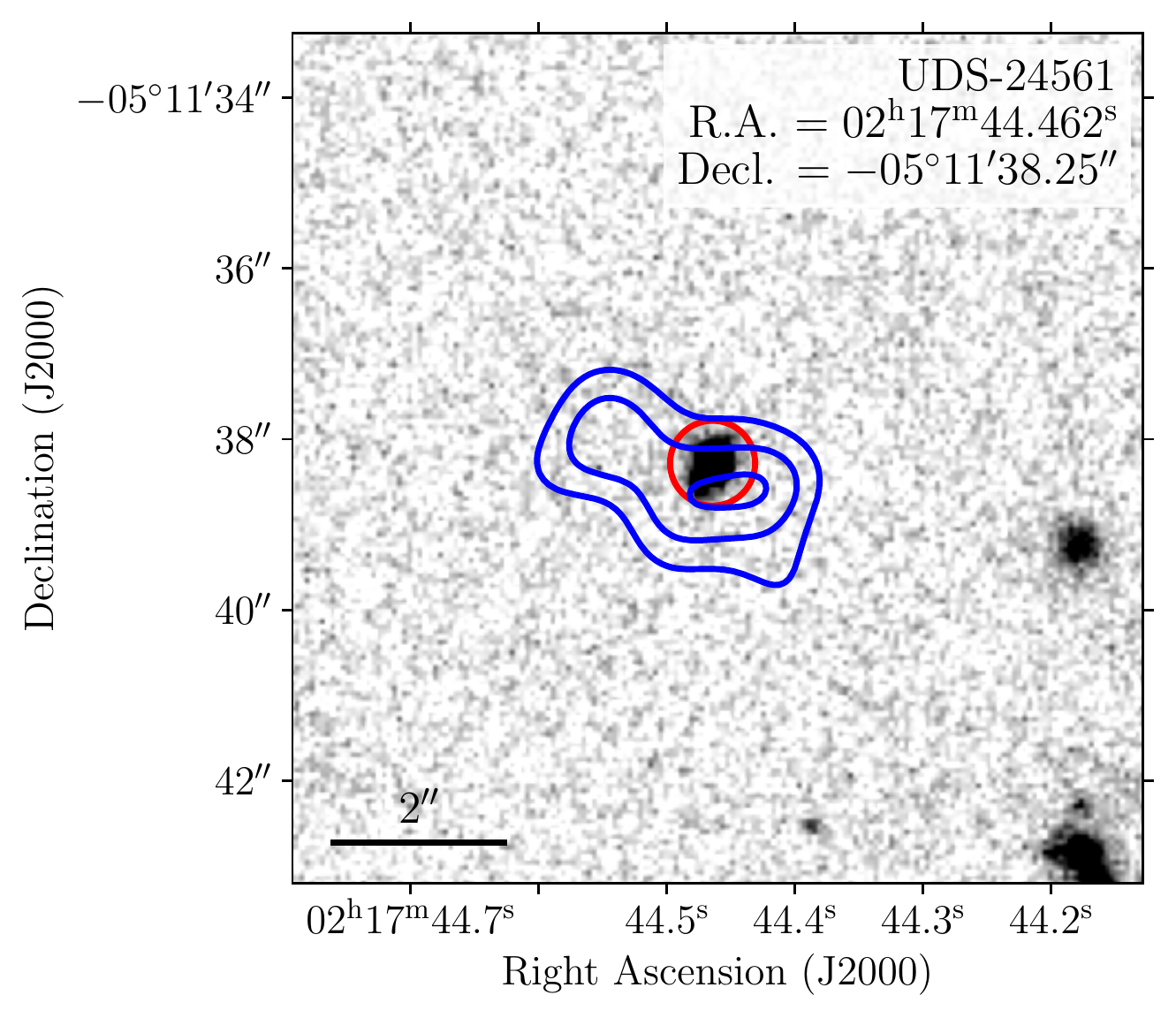}
\caption{Rest-frame UV and X-ray imaging of UDS-24561. The black-white image shows the {\it HST} F814W (rest-frame UV) postage stamp ($10^{\prime\prime}\times10^{\prime\prime}$) of UDS-24561. The red circle shows the region centered on UDS-24561 with a radius of $0.5$~arcsec. The blue contour overlaying the F814W imaging shows the full band ($0.5-10$ keV) deep {\it Chandra} X-ray imaging, with contour lines represent count levels of $1$, $1.5$, and $2$. The {\it Chandra} X-ray source is consistent with the {\it HST} imaging, indicating that the X-ray emission comes from UDS-24561.}
\label{fig:image}
\end{center}
\end{figure}


\begin{figure}
\begin{center}
\includegraphics[width=\linewidth]{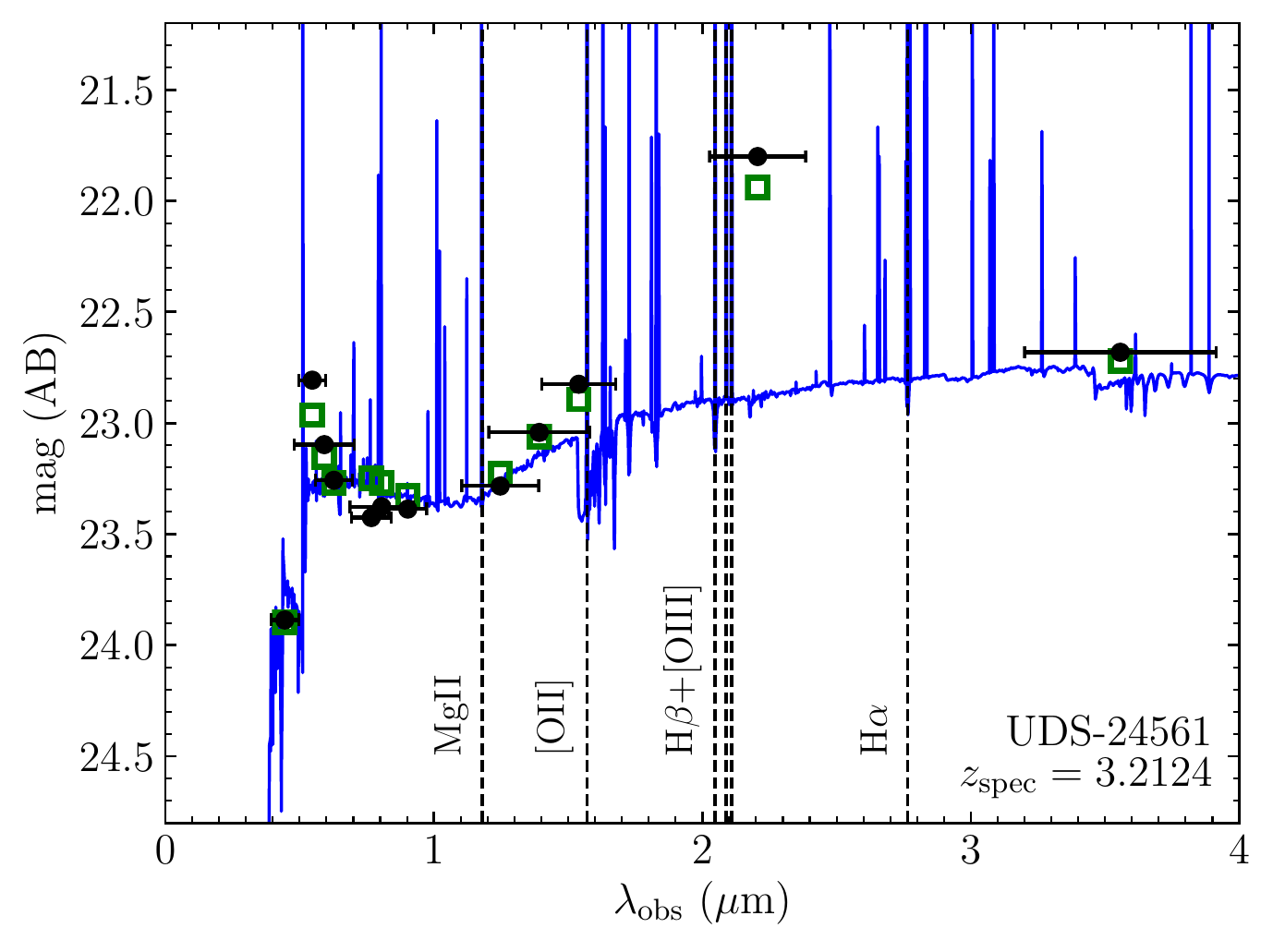}
\caption{The spectral energy distribution of UDS-24561. Observed broadband photometry extracted from the \citet{Skelton2014} catalogs is shown as solid black circles. The best-fit SED model inferred from the Bayesian spectral energy distribution modeling and interpreting tool BEAGLE \citep{Chevallard2016} are plotted by solid blue lines, and synthetic photometry is presented by open green squares. The SED of UDS-24561 shows a blue UV slope ($\beta=-2.34$) and an intense $K$-band flux excess indicating a large [O~{\scriptsize III}]+H$\beta$ EW ($\simeq1300$~\AA).}
\label{fig:sed}
\end{center}
\end{figure}


\section{Results} \label{sec:results}

The luminous X-ray detection of UDS-24561 reveals that this system hosts an AGN. In this section, we explore the spectral properties of UDS-24561 in more detail with the optical and near-infrared spectra described in Section~\ref{sec:observations}. We first present the emission line identification and measurements (Section~\ref{sec:line}), then compare the observed line ratios to line diagnostics developed from photoionisation models to examine their capability of identifying the nature of the ionising sources (Section~\ref{sec:diagnostics}).

\subsection{Emission line measurements} \label{sec:line}

\begin{figure*}
\begin{center}
\includegraphics[width=\linewidth]{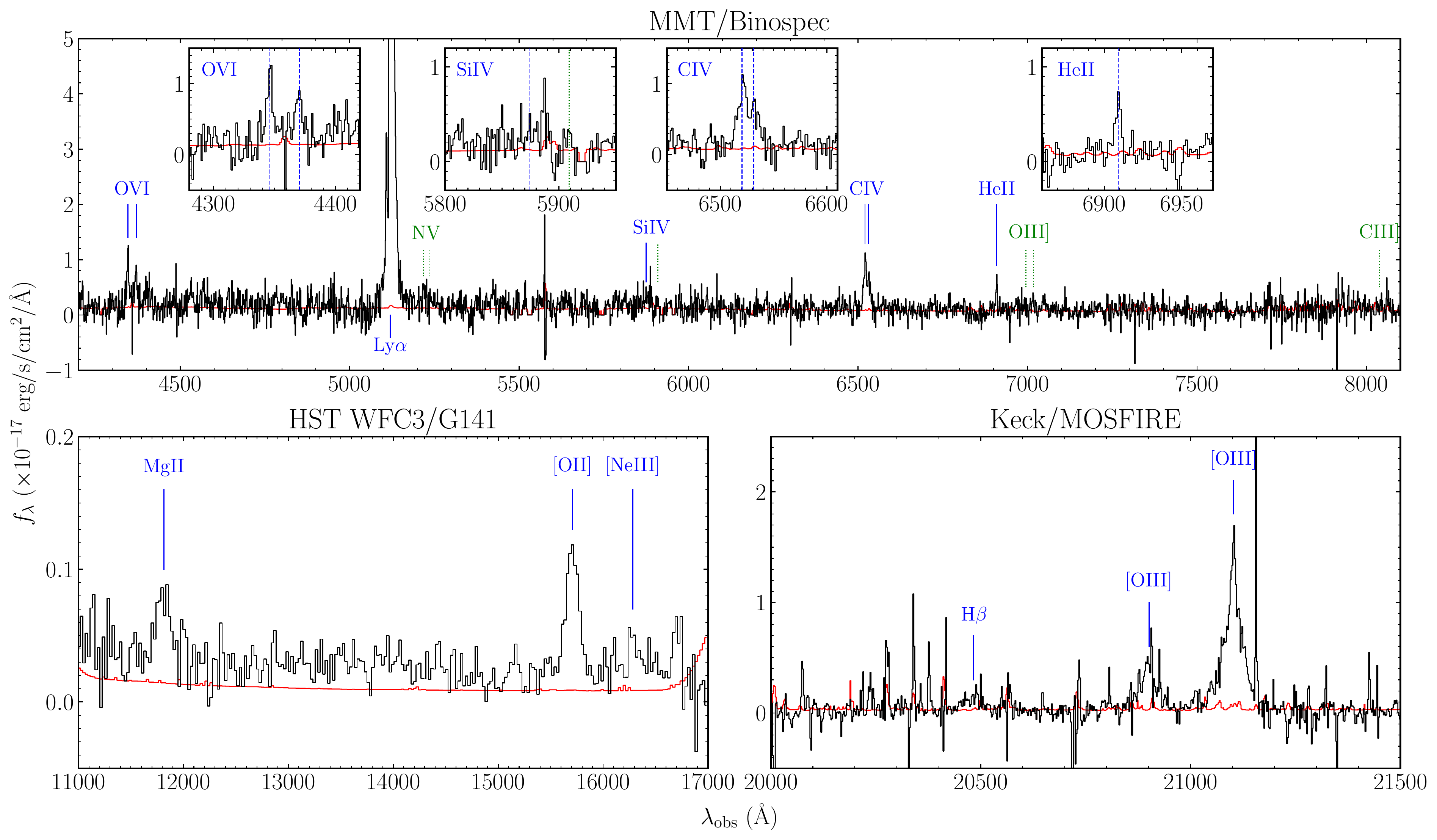}
\caption{Rest-frame UV and optical spectra of UDS-24561. MMT/Binospec, {\it HST} WFC3/G141, and Keck/MOSFIRE spectra are shown in the top, the bottom left, and the bottom right panel, respectively. The black and red solid histograms represent the observed flux and $1\sigma$ uncertainty. Detected emission lines including O~{\scriptsize VI}~$\lambda\lambda1032,1038$, Ly$\alpha$, Si~{\scriptsize IV}~$\lambda1393$ (tentatively), C~{\scriptsize IV}~$\lambda\lambda1548,1550$, He~{\scriptsize II}~$\lambda1640$, blended Mg~{\scriptsize II}~$\lambda\lambda2976,2803$, blended [O~{\scriptsize II}]~$\lambda\lambda3727,3729$, [Ne~{\scriptsize III}]~$\lambda3869$, H$\beta$, [O~{\scriptsize III}]~$\lambda4959$, and [O~{\scriptsize III}]~$\lambda5007$ are marked by blue solid lines. Predicted positions of non-detected N~{\scriptsize V}, Si~{\scriptsize IV}~$\lambda1402$, O~{\scriptsize III}], and C~{\scriptsize III}] lines are marked by green dotted lines. In the top panel, we also show the zoom-in spectra of O~{\scriptsize VI}, Si~{\scriptsize IV}, C~{\scriptsize IV}, and He~{\scriptsize II} emission lines.}
\label{fig:spectra}
\end{center}
\end{figure*}

The rest-frame UV and optical spectra of UDS-24561 are shown in Fig.~\ref{fig:spectra}. We visually inspect the spectra and identify a number of emission lines including O~{\small VI}~$\lambda\lambda1032,1038$, Ly$\alpha$, C~{\small IV}~$\lambda\lambda1548,1550$, He~{\small II}~$\lambda1640$, Mg~{\small II}~$\lambda\lambda2976,2803$, [O~{\small II}]~$\lambda\lambda3727,3729$, [Ne~{\small III}]~$\lambda3869$, H$\beta$, [O~{\small III}]~$\lambda4959$, [O~{\small III}]~$\lambda5007$, and tentatively Si~$\lambda1393$ ($\sim2\sigma$), while N~{\small V}~$\lambda\lambda1238,1243$, Si~$\lambda1402$, O~{\small III}]~$\lambda\lambda1661,1666$, and C~{\small III}]~$\lambda\lambda1907,1909$ emission lines are not detected. The systemic redshift of UDS-24561 is determined using the line centers of non-resonant emission lines with spectral resolution $R>1000$, i.e., He~{\small II}~$\lambda1640$, [O~{\small III}]~$\lambda4959$, and [O~{\small III}]~$\lambda5007$ emission lines, resulting in $z_{\rm{sys}}=3.2124\pm0.0006$. 

Emission line fluxes of UDS-24561 are determined from the extracted 1D spectra. For He~{\small II}~$\lambda1640$, we fit the line profile with a single Gaussian to derive the line flux. For O~{\small VI}~$\lambda\lambda1032,1038$ and C~{\small IV}~$\lambda\lambda1548,1550$ doublets, we fit two Gaussians to derive the line flux of each individual component of the doublet. For [O~{\small III}]~$\lambda5007$, we fit two Gaussians with a broad and a narrow component since the emission line cannot be well fitted by a single Gaussian, and the narrow component was used to compute the systemic redshift. The Ly$\alpha$ emission line of UDS-24561 shows a double-peak profile and we fit four Gaussians, so that both  blue and red Ly$\alpha$ components can be fitted by broad and a narrow components. For H$\beta$ and [O~{\small III}]~$\lambda4959$, we calculate the line flux using direct integration instead of fitting Gaussians since these two emission lines are contaminated by sky lines and thus have low signal-to-noise ratio (S/N $<4$). We also use direct integration to compute the flux of the tentative Si~$\lambda1393$ emission line (S/N $=2$). For the blended Mg~{\small II}~$\lambda\lambda2976,2803$, [O~{\small II}]~$\lambda\lambda3727,3729$ doublets and [Ne~{\small III}]~$\lambda3869$ which lie in the {\it HST} grism spectrum, we use the line flux provided in the \citet{Momcheva2016} catalogs. Previous studies show that the line fluxes measured from {\it HST} grism spectra are consistent with the values measured from ground-based observations (e.g., \citealt{Kriek2015}; \citetalias{Tang2019}). For the non-detected N~{\small V}~$\lambda\lambda1238,1243$, Si~$\lambda1402$, O~{\small III}]~$\lambda\lambda1661,1666$, and C~{\small III}]~$\lambda\lambda1907,1909$ lines, we derive $3\sigma$ upper limits by summing the error spectrum in quadrature over $\sim200$~km$/$s, a value consistent with the upper bound of line widths found for UV metal lines \citep[e.g.,][]{Stark2014}. 

We next compute the corresponding rest-frame EW for each emission line. Robust measurements of continuum flux are required to compute EWs. For emission lines (except for Ly$\alpha$) lying in the MMT/Binospec spectrum which shows bright continuum emission (S/N $>5$), we derive the continuum flux density in a clean window of $\pm150$~\AA\ near the emission line in the extracted 1D spectrum. For Ly$\alpha$ we estimate the continuum flux by averaging the flux over rest-frame $1270-1300$~\AA\ in order to avoid contamination from nearby features (H~{\small I} absorption blueward of Ly$\alpha$, and N~{\small V}, Si~{\small II} redward of Ly$\alpha$; e.g., \citealt{Matthee2021}). For Mg~{\small II}, [O~{\small II}], and [Ne~{\small III}] in the {\it HST} grism spectrum, we use the EWs provided in the \citet{Momcheva2016} catalogs. We also compute these EWs by measuring the continuum flux from the grism spectrum and they are in agreement with the EWs provided by the 3D-HST team. For H$\beta$ and [O~{\small III}] in the Keck/MOSFIRE spectrum where a bright continuum emission is not available, we derive the continuum flux by subtracting the emission line fluxes from the total $K$-band flux. We summarise the emission line measurements of UDS-24561 in Table~\ref{tab:line}, including the deconvolved FWHMs (after subtracting the instrument resolution in quadrature), line fluxes, and EWs. 

The most prominent emission line detected in the spectra of UDS-24561 is [O~{\small III}]~$\lambda5007$, with EW $=935\pm167$~\AA. The total [O~{\small III}]+H$\beta$ EW measured from spectrum is $1298\pm193$~\AA, which is consistent with the value inferred from the $K$-band flux excess (EW $\simeq1300$~\AA; Fig.~\ref{fig:sed}). Such intense optical line emission is extremely rare at $z\sim0-3$, but becomes more common in the reionisation era (e.g., $20$~per~cent of the $z\simeq7$ population has [O~{\small III}]+H$\beta$ EW $>1200$~\AA; \citealt{Endsley2021a}). 

\begin{figure}
\begin{center}
\includegraphics[width=\linewidth]{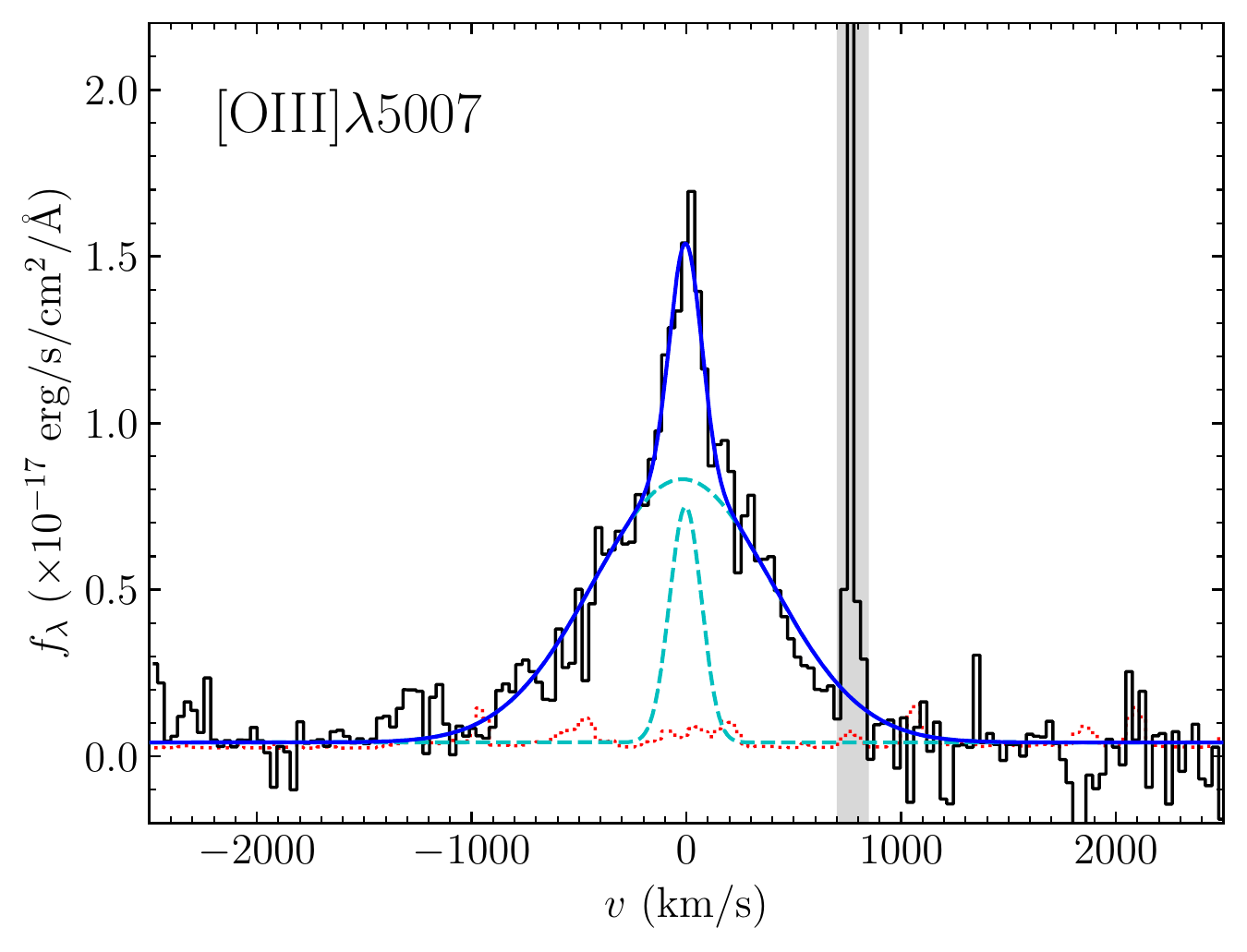}
\caption{Observed [O~{\scriptsize III}]~$\lambda5007$ emission line of UDS-24561 and the best-fit line profile. The x-axis shows the velocity offset with respect to the systemic redshift. Observed flux and $1\sigma$ uncertainty are shown by the black solid histogram and the red dotted line. The grey shaded region marks the region that is strongly contaminated by sky line residues. The [O~{\scriptsize III}]~$\lambda5007$ emission line is best fitted by a double-Gaussian profile (blue solid line), with a narrow component with FWHM $=154\pm21$~km$/$s and a broad component with FWHM $=977\pm131$~km$/$s (both shown by the cyan dashed lines).}
\label{fig:oiii}
\end{center}
\end{figure}

The [O~{\small III}] doublets and H$\beta$ emission lines show extended profiles (Fig.~\ref{fig:spectra}), and the [O~{\small III}]~$\lambda5007$ line is best fitted by two Gaussians: a narrow component with FWHM $=154\pm21$~km$/$s and a broad component with FWHM $=977\pm131$~km$/$s (Fig.~\ref{fig:oiii}). A broad line could be due to merger activity, emission from the broad line region of an AGN or AGN-driven outflows. We note that the {\it HST} image of UDS-24561 (Fig.~\ref{fig:image}) does not indicate any evidence of merger activity. Moreover, the broad [O~{\small III}] line width of UDS-24561 is somewhat less than the typical line width seen in AGN broad line regions (FWHM $\gtrsim2000$~km$/$s). Although supernova feedback or shock-ionised outflows can also produce broad line emission, the FWHMs ($\sim300-600$~km$/$s; e.g., \citealt{Veilleux2005,Freeman2019,Matthee2021}) are smaller than that of UDS-24561. On the other hand, the broad optical line velocity of UDS-24561 is in agreement with those of AGN outflows \citep[e.g.,][]{ForsterSchreiber2014,Genzel2014,Leung2017,Leung2019}. We estimate a lower limit to the broad [O~{\small III}]~$\lambda5007/\rm{H}\beta$ ratio using the total flux of H$\beta$ (since we cannot accurately decouple the broad and narrow emission components for H$\beta$). The line ratio ($>9.3$ at $3\sigma$) is consistent with those of AGN \citep[e.g.,][]{Coil2015,Azadi2017} and greater than the maximum ratio predicted by starburst models \citep[e.g.,][]{Kauffmann2003,Kewley2013}. Thus we conclude that the broad optical lines of UDS-24561 likely originate from an AGN-driven outflow.

\begin{figure}
\begin{center}
\includegraphics[width=\linewidth]{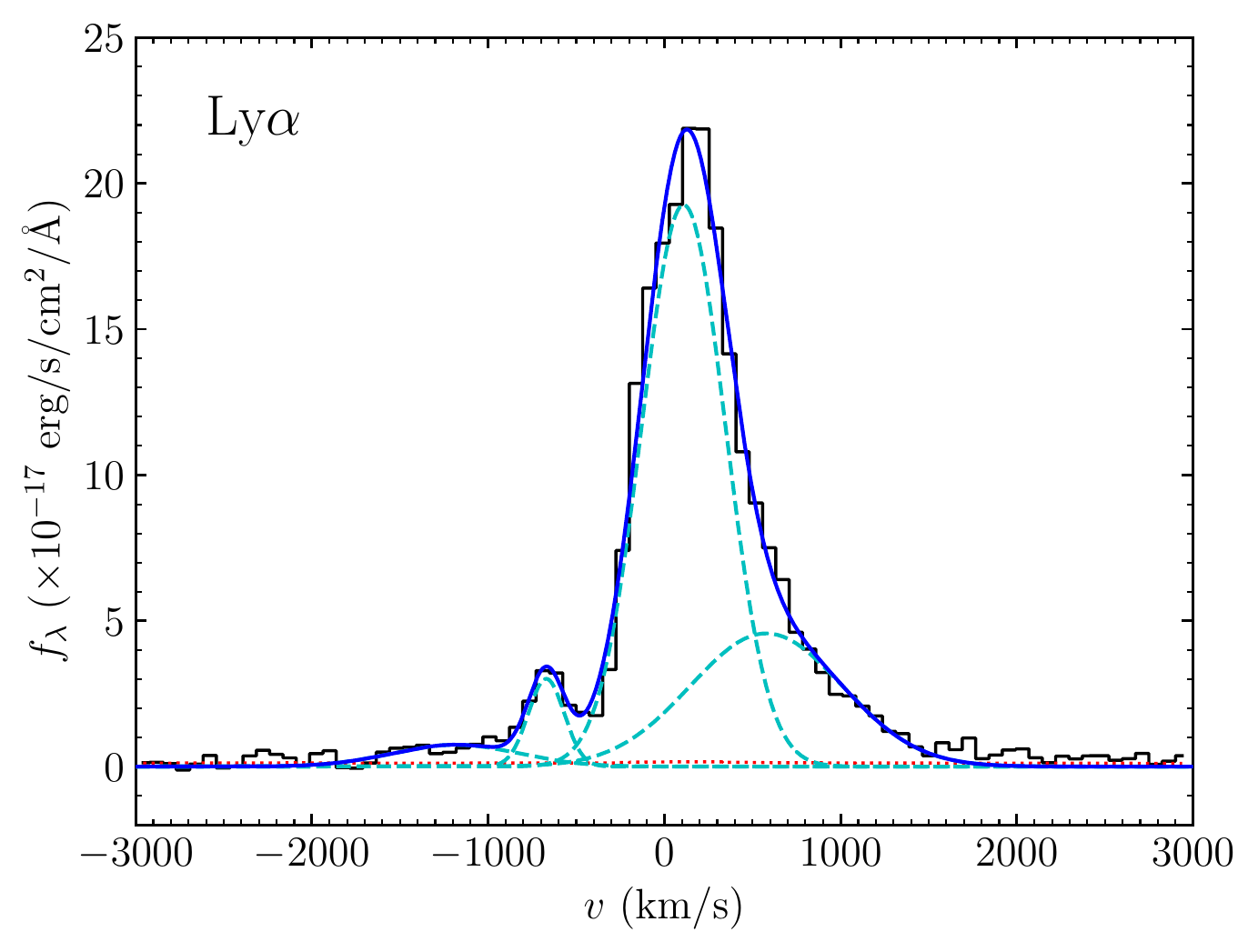}
\caption{Observed Ly$\alpha$ emission line of UDS-24561 and the best-fit line profile. The x-axis shows the velocity offset with respect to the systemic redshift. Observed flux and $1\sigma$ uncertainty are shown by the black solid histogram and the red dotted line. The Ly$\alpha$ emission line shows a double-peaked profile, and each of the blue peak and red peak Ly$\alpha$ line is best fitted by a double-Gaussian profile (a quadruple-Gaussian profile for the total Ly$\alpha$ emission, shown by the blue solid line), with a narrow component (FWHM $=87-513$~km$/$s) and a broad component (FWHM $=814-987$~km$/$s). Each of the single Gaussian component is shown by the cyan dashed line. The blue (red) wing feature extends to a maximal velocity of $\sim-2000$~km$/$s ($+1600$~km$/$s).}
\label{fig:lya}
\end{center}
\end{figure}

The Ly$\alpha$ emission line of UDS-24561 shows a more complex profile (Fig.~\ref{fig:lya}). The emission is double-peaked, and both the blue and red Ly$\alpha$ components show extended emission line features. None of the blue and red Ly$\alpha$ component could be fitted by a single Gaussian. However, we find that each Ly$\alpha$ component is best fitted by two Gaussians including a narrow (FWHM $=87-513$~km$/$s) and a broad (FWHM $=814-987$~km$/$s) component, indicating the blue and red wings extending to maximal velocities of $-2000$~km$/$s and $+1600$~km$/$s (Fig.~\ref{fig:lya}). The broad wing features, together with the small velocity offset of red peak Ly$\alpha$ (which is also consistent with the center of the narrow red Ly$\alpha$ line) with respect to the systemic redshift ($\Delta v_{\rm{Ly}\alpha}=+141\pm76$~km$/$s), suggest conditions that are conducive to leaking Ly$\alpha$ photons \citep[e.g.,][]{Erb2014,Hashimoto2015,Henry2015,Martin2015}. The total Ly$\alpha$ luminosity and EW of UDS-24561 are extremely large, with $L_{\rm{Ly}\alpha}=2.8\pm0.3\times10^{44}$~erg~s$^{-1}$ and EW$_{\rm{Ly}\alpha}=436\pm45$~\AA, which the $L_{\rm{Ly}\alpha}$ is $\sim50$ times the typical Ly$\alpha$ luminosity of LAEs at $z\sim3$ ($L^*_{\rm{Ly}\alpha}\simeq10^{42.7}$~erg~s$^{-1}$; e.g., \citealt{Sobral2018a}). Such luminous $L_{\rm{Ly}\alpha}$ and $L_{\rm{UV}}$ ($\sim5$ times the typical UV luminosity at $z\sim3$; \citealt{Parsa2016}), together with the blue UV slope, are consistent with the picture that the physics of accretion discs of massive black holes can destroy the dust in very luminous systems and allow Ly$\alpha$ and UV photons to escape \citep[e.g.,][]{Sobral2018b}. 

Intense high ionisation emission lines from O~{\small VI},\footnote{The spectral resolution of MMT/Binospec allows us to deblend Ly$\beta$ (at rest-frame $1025.73$~\AA) and O~{\small VI}~$\lambda\lambda1032,1038$ lines}, Si~{\small IV}~$\lambda1393$, C~{\small IV}, and He~{\small II} have also been detected in the rest-frame UV spectrum of UDS-24561 (Fig.~\ref{fig:spectra}), suggesting a hard ionising radiation field powered by an AGN \citep[e.g.,][]{Feltre2016,Volonteri2017}. Unlike Ly$\alpha$, H$\beta$, or [O~{\small III}], we do not detect significant broad emission features for these high ionisation lines. As the UV line fluxes are relatively fainter, it could be more difficult to detect broad emission components and deeper spectra are required to measure such features. The line profile of He~{\small II} is narrow and the line width is comparable to the instrument resolution. We estimate the $3\sigma$ upper limit of the deconvolved FWHM $<216$~km$/$s for He~{\small II}, which is consistent with the FWHM of the narrow [O~{\small III}]~$\lambda5007$ emission line ($154\pm21$~km$/$s). Individual components of the O~{\small VI} and C~{\small IV} doublets show wider FWHMs ($=340-499$~km$/$s), which could be due to the resonant nature of these emission lines.

The O~{\small VI}, Si~{\small IV}~$\lambda1393$, C~{\small IV}, and He~{\small II} EWs derived from the spectrum of UDS-24561 are EW$_{\rm{OVI}}=16.2\pm2.6$~\AA, EW$_{\rm{SiIV}\lambda1393}=1.1\pm0.5$~\AA, EW$_{\rm{CIV}}=27.4\pm4.0$~\AA\ and EW$_{\rm{HeII}}=5.3\pm1.3$~\AA. Significant stellar or interstellar absorption features are not seen in the vicinity of O~{\small VI}, Si~{\small IV}~$\lambda1393$, or C~{\small IV} emission lines, although we cannot rule out a modest level of absorption and hence these emission line EWs could be even larger. The majority of metal-poor SFGs do not present such large EWs \citep[e.g.,][]{Erb2010,Vanzella2016,Senchyna2017,Senchyna2019,Berg2018,Berg2019,Du2020}, except a few sources show similar C~{\small IV} EWs \citep{Stark2015,Mainali2017,Vanzella2017}. On the contrary, the high ionisation line EWs of UDS-24561 are comparable to the values measured in AGN \citep[e.g.,][]{Hainline2011,LeFevre2019,Mignoli2019,Grazian2020,Saxena2020}. Using the available UV emission line ratios of UDS-24561, we will test whether the line diagnostics developed to identify ionising sources are capable of distinguishing between spectra powered by blue, extreme optical line emitting AGN and massive stars in Section~\ref{sec:diagnostics}.


\begin{table}
\begin{tabular}{|c|c|c|c|c|}
\hline
Line & $\lambda_{\rm{rest}}$ & FWHM & Flux & EW \\
 & (\AA) & (km$/$s) & ($10^{-17}$~erg~s$^{-1}$~cm$^{-2}$) & (\AA) \\
\hline
\hline
$\rm{O}$~{\scriptsize VI} & $1031.91$ & $354\pm54$ & $6.32\pm1.27$ & $8.6\pm1.7$ \\
... & $1037.61$ & $499\pm99$ & $5.56\pm1.46$ & $7.6\pm2.0$ \\
Ly$\alpha_{\rm{b,n}}$ & $1215.67^{\rm{a}}$ & $87\pm11$ & $13.10\pm2.17$ & $19\pm3$ \\
Ly$\alpha_{\rm{b,b}}$ & $1215.67^{\rm{b}}$ & $814\pm269$ & $11.45\pm4.20$ & $16\pm6$ \\
Ly$\alpha_{\rm{b,t}}$ & $1215.67^{\rm{c}}$ & ... & $24.55\pm4.72$ & $35\pm7$ \\
Ly$\alpha_{\rm{r,n}}$ & $1215.67^{\rm{d}}$ & $513\pm23$ & $195.8\pm21.8$ & $280\pm31$ \\
Ly$\alpha_{\rm{r,b}}$ & $1215.67^{\rm{e}}$ & $987\pm168$ & $84.07\pm21.78$ & $120\pm31$ \\
Ly$\alpha_{\rm{r,t}}$ & $1215.67^{\rm{f}}$ & ... & $279.9\pm30.8$ & $401\pm44$ \\
N~{\scriptsize V} & $1238.82$ & ... & $<1.46$ & $<2.1$ \\
... & $1242.80$ & ... & $<1.43$ & $<2.0$ \\
Si~{\scriptsize IV} & $1393.76$ & ... & $0.74\pm0.34$ & $1.1\pm0.5$ \\
... & $1402.77$ & ... & $<1.43$ & $<2.1$ \\
C~{\scriptsize IV} & $1548.19$ & $440\pm65$ & $9.74\pm1.62$ & $18.2\pm3.0$ \\
... & $1550.77$ & $340\pm83$ & $4.93\pm1.39$ & $9.2\pm2.6$ \\
He~{\scriptsize II} & $1640.42$ & $<216$ & $2.60\pm0.64$ & $5.3\pm1.3$ \\
$\rm{O}$~{\scriptsize III}$]$ & $1660.81$ & ... & $<1.15$ & $<2.3$ \\
... & $1666.15$ & ... & $<1.07$ & $<2.2$ \\
C~{\scriptsize III}$]$ & $1908^{\rm{g}}$ & ... & $<3.30$ & $<9.7$ \\
Mg~{\scriptsize II} & $2798^{\rm{h}}$ & ... & $8.36\pm1.18$ & $63\pm10$ \\
$[\rm{O}$~{\scriptsize II}$]$ & $3728^{\rm{i}}$ & ... & $14.94\pm0.65$ & $172\pm12$ \\
$[$Ne~{\scriptsize III}$]$ & $3870.16$ & ... & $3.12\pm0.70$ & $27\pm7$ \\
H$\beta$ & $4862.69$ & ... & $4.76\pm2.08$ & $69\pm30$ \\
$[\rm{O}$~{\scriptsize III}$]$ & $4960.30$ & ... & $21.51\pm6.66$ & $294\pm91$ \\
$[\rm{O}$~{\scriptsize III}$]_{\rm{n}}$ & $5008.24^{\rm{j}}$ & $154\pm21$ & $9.28\pm1.90$ & $129\pm26$ \\
$[\rm{O}$~{\scriptsize III}$]_{\rm{b}}$ & $5008.24^{\rm{k}}$ & $977\pm131$ & $58.05\pm11.90$ & $805\pm165$ \\
$[\rm{O}$~{\scriptsize III}$]_{\rm{t}}$ & $5008.24^{\rm{l}}$ & ... & $67.38\pm12.05$ & $935\pm167$ \\
\hline
\end{tabular}
\caption{Rest-frame UV and optical emission line measurements of UDS-24561 at $z=3.2124$, including the deconvolved FWHM (Column 3), line flux (Column 4), and rest-frame EW (Column 5). For non-detected N~{\scriptsize V}, O~{\scriptsize III}], and the blended C~{\scriptsize III}] lines, $3\sigma$ upper limits are provided. \newline 
$^{\rm{a}}$ Narrow component of the blue peak of Ly$\alpha$. \newline 
$^{\rm{b}}$ Broad component of the blue peak of Ly$\alpha$. \newline 
$^{\rm{c}}$ Total of the blue peak of Ly$\alpha$. \newline 
$^{\rm{d}}$ Narrow component of the red peak of Ly$\alpha$. \newline 
$^{\rm{e}}$ Broad component of the red peak of Ly$\alpha$. \newline 
$^{\rm{f}}$ Total of the red peak of Ly$\alpha$. \newline 
$^{\rm{g}}$ Total of the blended C~{\scriptsize III}]~$\lambda\lambda1907,1909$. \newline 
$^{\rm{h}}$ Total of the blended Mg~{\scriptsize II}~$\lambda\lambda2796,2803$. \newline 
$^{\rm{i}}$ Total of the blended [O~{\scriptsize II}]~$\lambda\lambda3727,3729$. \newline 
$^{\rm{j}}$ Narrow component of [O~{\scriptsize III}]~$\lambda5007$. \newline 
$^{\rm{k}}$ Broad component of [O~{\scriptsize III}]~$\lambda5007$. \newline 
$^{\rm{l}}$ Total of [O~{\scriptsize III}]~$\lambda5007$.}
\label{tab:line}
\end{table}


\subsection{Emission line diagnostics} \label{sec:diagnostics}

Various diagnostics involving rest-frame UV emission lines have been developed to determine the nature of the ionising sources (i.e., star formation, nuclear activity, or shocks) using photoionisation models \citep[e.g.,][]{Feltre2016,Nakajima2018,Hirschmann2019}. The emission line measurements of UDS-24561 provide a unique opportunity to examine the capability of the line diagnostics in identifying AGN in extreme optical line emitting systems with blue UV slopes. Since blue, extreme [O~{\small III}] emitting sources are common at $z>7$ \citep[e.g.,][]{Smit2015,Endsley2021a}, and UV emission lines will likely remain the most useful diagnostics for studies of the most distant galaxies, our present analysis will provide useful insight into identifying AGN in the reionisation era.

For emission line ratios predicted from photoionisation models, we consider the SFG models taken from \citet{Gutkin2016} and the latest version of \citet{Feltre2016} AGN narrow line region models presented in \citet{Mignoli2019}. The nebular emission of both models is computed using the photoionisation code {\footnotesize CLOUDY} \citep{Ferland2013}. In this work, we focus on emission line ratios involving O~{\small III}]~$\lambda\lambda1661,1666$ and C~{\small IV}~$\lambda\lambda1548,1550$, He~{\small II}~$\lambda1640$. Since the ionising spectra of metal-poor hot stars decline rapidly around $50$~eV while those of AGN extend to much higher energies (e.g., Figure 1 in \citealt{Feltre2016}), the oxygen atoms in AGN are expected to be largely triply ionised while C~{\small IV} and He~{\small II} emission is strong compared to SFGs. We thus expect a decrease in O~{\small III}] to He~{\small II} or C~{\small IV} ratio in AGN, and these line ratios can be used to determine the ionising sources. 

\begin{figure*}
\begin{center}
\includegraphics[width=\linewidth]{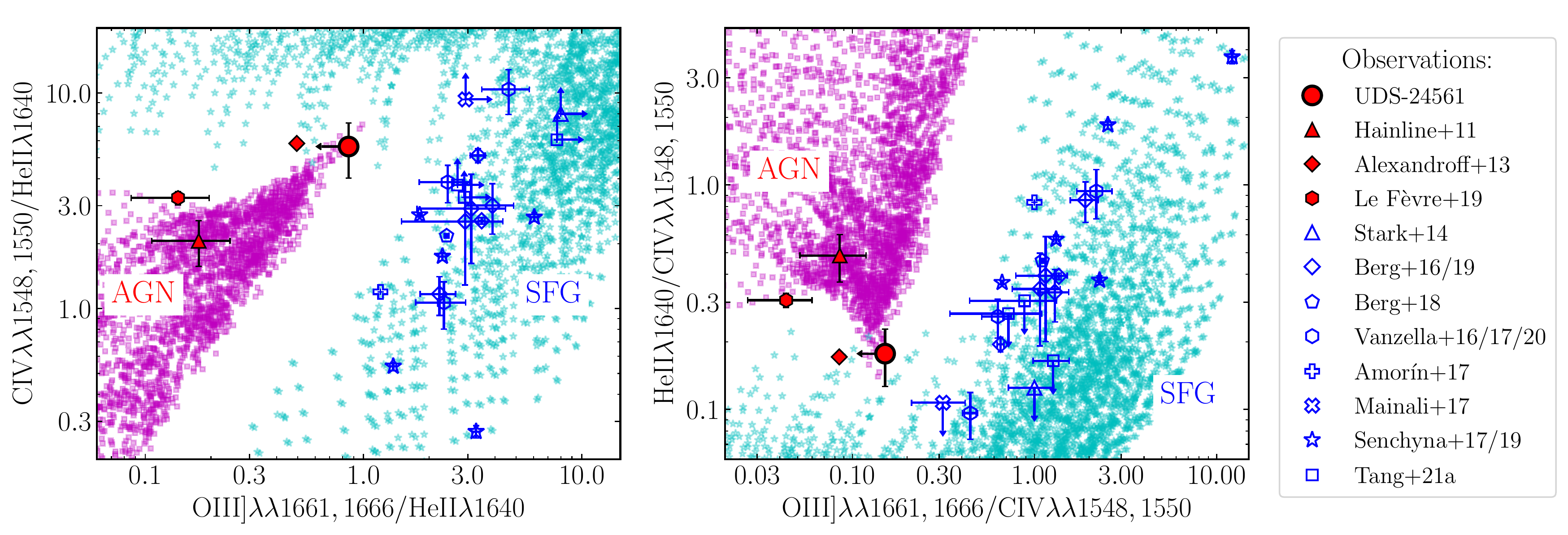}
\caption{{\it Left:} C~{\scriptsize IV}~$\lambda\lambda1548,1550/$He~{\scriptsize II}~$\lambda1640$ versus O~{\scriptsize III}]~$\lambda\lambda1661,1666/$He~{\scriptsize II}~$\lambda1640$ diagnostic diagram \citep{Feltre2016}. {\it Right:} He~{\scriptsize II}~$\lambda1640/$C~{\scriptsize IV}~$\lambda\lambda1548,1550$ versus O~{\scriptsize III}]~$\lambda\lambda1661,1666/$C~{\scriptsize IV}~$\lambda\lambda1548,1550$ diagnostic diagram \citep{Mainali2017}. Line ratios taken from photoionisation models are plotted as magenta squares for AGN models \citep{Feltre2016,Mignoli2019} and cyan stars for SFG models \citep{Gutkin2016}. Observed data of UDS-24561 are shown by red circles in both diagnostic diagrams. From the literature, we overlay the data of AGN composite spectra (\citealt{Hainline2011,Alexandroff2013,LeFevre2019}; red solid symbols) and SFGs (\citealt{Stark2014,Berg2016,Berg2018,Berg2019,Vanzella2016,Vanzella2017,Vanzella2020,Amorin2017,Mainali2017,Senchyna2017,Senchyna2019}; \citetalias{Tang2021a}; blue open symbols) at $z=0-7$ in both diagrams. Compared to SFGs, AGN are characterised by smaller O~{\scriptsize III}]$/$He~{\scriptsize II} and O~{\scriptsize III}]$/$C~{\scriptsize IV} ratios owing to the hard ionising spectra that power strong He~{\scriptsize II} and C~{\scriptsize IV} emission and triply ionise oxygen.}
\label{fig:diagnostic}
\end{center}
\end{figure*}

In the left panel of Fig.~\ref{fig:diagnostic}, we show the C~{\small IV}$/$He~{\small II} versus O~{\small III}]$/$He~{\small II} diagnostic for both nebular emission from photoionisation models (magenta squares: AGN; cyan stars: SFG) and observations (red solid symbols: AGN; blue open symbols: SFG). Photoionisation models and observations at $z\simeq0-4$ in literature indicated that narrow-line AGN reach smaller C~{\small IV}$/$He~{\small II} ratios ($<7$) as well as smaller O~{\small III}]$/$He~{\small II} ratios at fixed C~{\small IV}$/$He~{\small II} compared to non-AGN SFGs. The line ratios of the blue, extreme [O~{\small III}] emitting AGN, UDS-24561, are also consistent with this picture. Using the AGN models in \citet{Feltre2016}, we find that the UV line ratios of UDS-24561 are best reproduced by an ionising spectrum with a slope of $\alpha\simeq-2.0$ (assuming $f_{\nu}\propto\nu^{\alpha}$ at rest-frame wavelength $\lambda<912$~\AA) and a very large ionisation parameter ($\log{U}\simeq-1.0$). We also notice that the line ratios of UDS-24561 are located in the AGN $-$ SFG region, implying that they can be reproduced by a few SFG models as well. However, for non-AGN SFG models to reproduce the line ratios of UDS-24561, much lower ionisation parameters ($\log{U}<-3.0$) and larger gas-phase metallicities ($Z>Z_{\odot}$) are required, both conditions allowing to reduce the emissivity of O~{\small III}] relative to He~{\small II}. SFG models with such conditions are not able to reproduce the large C~{\small IV} EWs seen in AGN.\footnote{Using the BEAGLE tool \citep{Chevallard2016} which adopted the latest version of stellar population synthesis code in \citet{Bruzual2003} and the nebular emission models from \citet{Gutkin2016}, the largest predicted C~{\scriptsize IV} EW for models with $\log{U}<-3.0$ and $Z>Z_{\odot}$ is only $1$~\AA\ assuming a constant star formation history, which is significantly smaller than the C~{\scriptsize IV} EWs of AGN.} 

The flux ratio of O~{\small III}] and C~{\small IV} provides another way to determine the ionising sources \citep[e.g.,][]{Mainali2017}. In the right panel of Fig.~\ref{fig:diagnostic} we show the He~{\small II}$/$C~{\small IV} versus O~{\small III}]$/$C~{\small IV} diagnostic. The O~{\small III}]$/$C~{\small IV} ratio of UDS-24561 is consistent with those of UV-selected AGN at similar redshift in literature \citep{Hainline2011,Alexandroff2013,LeFevre2019}, which are larger than the ratios of non-AGN metal-poor SFGs at fixed He~{\small II}$/$C~{\small IV} ratio. We also notice that the UV-blue AGN UDS-24561 appears to have smaller He~{\small II}$/$C~{\small IV} ratio compared to AGN with red UV continua at similar redshift in \citet{Hainline2011} and \citet{LeFevre2019}. This may suggest different properties between these two populations (e.g., metallicity or dust content of the host galaxies, see Section~\ref{sec:introduction}). The results presented in Fig.~\ref{fig:diagnostic} demonstrate that emission line diagnostics involving high ionisation lines such as C~{\small IV}, He~{\small II}, and O~{\small III}] are applicable to identifying AGN in extreme optical line emitting galaxies with blue UV continua. Future spectroscopic observations of a larger statistical sample of such systems will help to confirm the capability of the UV diagnostics in determining the ionising sources of reionisation-era galaxies.


\section{Discussion} \label{sec:discussion}

In the foregoing, we have demonstrated that, both from its detailed spectrum and X-ray luminosity, the extreme line emitting galaxy UDS-24561 contains an AGN. Yet it is also a close analogue of sources known to be common in the reionisation era. UDS-24561 displays a very blue UV slope ($\beta=-2.34$) and a large [O~{\small III}]+H$\beta$ EW ($=1298$~\AA). Rest-frame optical line emission appears to be prominent at $z\gtrsim7$ \citep[e.g.,][]{Labbe2013,DeBarros2019,Endsley2021a} and $20$~per~cent of this population has extremely large [O~{\small III}]+H$\beta$ EW of $>1200$~\AA\ \citep{Endsley2021a}, similar to the [O~{\small III}]+H$\beta$ EW of UDS-24561. The UV continuum slopes of the reionisation-era population ($\beta\simeq-2.0$) are also bluer than those of star-forming galaxies at lower redshifts \citep[e.g.,][]{Bouwens2012,Bouwens2014,Finkelstein2012}. In this section, we consider the implications of UDS-24561 for identifying AGN activity in reionisation-era galaxies and their contribution to the ionising background at $z>6$. 

UDS-24561 provides a detailed glimpse of a population of AGN that may become common in the reionisation era. Its distinct properties relative to more typical AGN at intermediate redshifts may be useful in locating AGN at very high redshift ($z>6$), leading to a better understanding of their role in contributing to cosmic reionisation. This is further supported by the fact that extreme optical line emitting systems are likely effective ionising agents (e.g., \citealt{Chevallard2018,Fletcher2019}; \citetalias{Tang2019,Tang2021b}). 

Unfortunately, identifying AGN activity at $z\gtrsim7$ with traditional means is likely to be difficult with current facilities. Current deep X-ray surveys do not have the sensitivity to locate individual examples. Consider, for example, an X-ray source with the same luminosity as UDS-24561. The full  ($0.5-10$~keV), soft ($0.5-2$~keV), and hard band ($2-10$~keV) fluxes of UDS-24561 at $z=3.2124$ are $3.01\times10^{-15}$, $0.59\times10^{-15}$, and $2.69\times10^{-15}$~erg~s$^{-1}$~cm$^{-2}$ \citep{Kocevski2018}, respectively. At $z=7$ such a source would have X-ray fluxes of $=5.84\times10^{-16}$, $1.14\times10^{-16}$, and $5.22\times10^{-16}$~erg~s$^{-1}$~cm$^{-2}$, respectively, in the full, soft, and hard bands assuming a power-law index of $\Gamma=1.7$ \citep{Brightman2014}. These fluxes are comparable to the $1\sigma$ sensitivity of the $600$~ks deep {\it Chandra} X-ray survey in the CANDELS/UDS field ($4.4\times10^{-16}$, $1.4\times10^{-16}$, and $6.5\times10^{-16}$~erg~s$^{-1}$~cm$^{-2}$ in the full, soft, and hard band; \citealt{Kocevski2018}). Although the $4$~Ms {\it Chandra} Deep Field-South survey reaches deeper flux limits \citep{Xue2011}, an individual AGN at $z=7$ with $L_{2-10\ \rm{keV}}\lesssim6\times10^{43}$~erg~s$^{-1}$ would still be undetected. Not surprisingly, no X-ray counterparts have been seen in all the six luminous ($L_{\rm{UV}}=1.3-2.5\times L^*_{\rm{UV}}$)  N~{\small V}-emitting galaxies at $z=6.5-9$ \citep{Tilvi2016,Hu2017,Laporte2017,Mainali2018,Sobral2019,Endsley2021b}. Future X-ray facilities will help accumulate larger AGN samples in the reionisation era \citep[e.g.,][]{Vito2018}. 

Likewise the broad H$\beta$ and [O~{\small III}] emission lines of UDS-24561, which likely reflect AGN-driven outflows, cannot readily be revealed in $z\gtrsim7$ spectra with current facilities. Detecting broad emission in individual lines requires relatively high resolution spectra ($R\gtrsim1000$) of good quality (S/N~$>5$). Although the blended [O~{\small III}]+H$\beta$ luminosities and EWs of $z\sim7-9$ systems can be estimated from the {\it Spitzer}/IRAC $[3.6]-[4.5]$ colors \citep[e.g.,][]{Labbe2013,Smit2014,DeBarros2019,Endsley2021a}, clearly broad emission cannot be discerned. As with studies of the diagnostic line ratios of [N~{\small II}]$/$H$\alpha$ or [S~{\small II}]$/$H$\alpha$ in the mid-infrared, rest-frame optical diagnostics of AGN activity at $z\gtrsim7$ must await spectroscopy with {\it JWST}/NIRSpec. 

We then consider the feasibility of detecting broad [O~{\small III}]~$\lambda5007$ emission in the reionisation era with {\it JWST}/NIRSpec. To robustly measure the broad feature, a good quality (S/N~$>5$) detection of the broad wings is required. If we consider a galaxy at $z=7$ harbouring AGN and with broad [O~{\small III}]~$\lambda5007$ EW $=800$~\AA\ and FWHM $=1000$~km$/$s (i.e., with parameters similar to those of UDS-24561), the predicted broad [O~{\small III}]~$\lambda5007$ emission line fluxes will be $4.3\times10^{-17}$ and $6.9\times10^{-18}$~erg~s$^{-1}$~cm$^{-2}$ for a source with the rest-frame optical continuum apparent magnitude of $25$ and $27$ ($-22$ and $-20$ in absolute magnitude; assuming a flat continuum in $f_{\nu}$), respectively. Using the medium resolution (R $=1000$) NIRSpec MSA observations with the G395M/F290LP disperser-filter combination, the {\it JWST} exposure time calculator predicts the broad wings at $\pm1000$~km$/$s away from the line center can be detected with S/N $=5$ in $1.5$ (continuum magnitude of $25$) and $9$ hours (continuum magnitude of $27$). The results demonstrates that reionisation-era AGN can be identified in a relatively short integration time with {\it JWST}/NIRSpec if they contain broad [O~{\small III}]~$\lambda5007$ emission like in UDS-24561.

Signs of AGN activity in the reionisation era can also be probed by rest-frame UV spectroscopy. The presence of high ionisation emission lines such as O~{\small VI} or N~{\small V}, and strong Mg~{\small II} line emission suggest hard radiation fields that are likely powered by AGN. As discussed in Section~\ref{sec:diagnostics} and earlier work \citep[e.g.,][]{Alexandroff2013,Feltre2016,Laporte2017,Mainali2017,Nakajima2018,Hirschmann2019,Mignoli2019}, UV line diagnostics are capable of characterising the shape of the ionising spectrum and hence distinguishing between AGN and star formation. However, this requires relatively high signal-to-noise ratio spectroscopy that can place key constraints on multiple high ionisation UV lines including C~{\small IV} and He~{\small II}. 

We have argued in this paper that UDS-24561 may provide a possible template for estimating the abundance of $z\gtrsim7$ galaxies hosting AGN if its combination of intense [O~{\small III}] emission and blue UV continuum slopes is a key characteristic. Although simply a hypothesis at present given the paucity of examples, the similarity between the properties of UDS-24561 and candidate galaxies with AGN in the reionisation era is striking.

Recently, \citet{Naidu2020} interpreted a rapid conclusion of reionisation over $6<z<7.5$ in the context of an additional population of rarer, more massive galaxies. In their Model I, bright galaxies with $M_{\rm{UV}}<18$, which account for less than $5$~per~cent of the early population, are arranged to account for $>50$~per~cent of the reionising photon budget. Via this hypothetical model, the rapid evolution in the redshift-dependent neutral fraction $x_{\rm{HI}}$ in the IGM can be matched. Although clearly not an unique interpretation of the $x_{\rm{HI}}$ data, a possible physical explanation of such a mass-dependent contribution of ionising photons would be the late development of AGN in a subset of massive galaxies, each of which would have more powerful ionising capabilities. 

As an illustration of this hypothesis in the context of UDS-24561, we can estimate the additional contribution of ionising photons arising if some proportion of sources with intense [O~{\small III}] emission and blue UV continuum slopes have a $100$~per~cent escape fraction due to a powerful AGN. The assumed $100$~per~cent escape fraction for AGN is often adopted in studies modeling the contribution of AGN to the ionising background \citep[e.g.][]{Madau2015}. Recent works indicate slightly lower escape fraction for AGN \citep[e.g.][]{Romano2019,Iwata2021,Trebitsch2021}, with an average value of $f_{\rm{esc}}\sim0.75$ inferred from observations of bright quasars and faint AGN at $z\sim4$ \citep{Cristiani2016,Grazian2018}. If, for example, $20$~per~cent of such luminous galaxies harbour AGN and the escape fraction assigned to this subset is $f_{\rm{esc}}=1$ ($f_{\rm{esc}}=0.75$), and the remaining sources have $f_{\rm{esc}}=0.2$, AGN can readily supply the required energy budget of $50-60$~per~cent ($40-50$~per~cent) adopted in Model I in \citet{Naidu2020}. Indeed, there is room to relax the adopted escape fraction for non-AGN sources to lower values, depending on the adopted luminosity limit for hosting AGN.

Although our adopted AGN fraction of $20$~per~cent in luminous $z\simeq6-7$ is arbitrary, we note that \citet{Endsley2021a} find $20$~per~cent of $z\simeq7$ sources have intense [O~{\small III}]+H$\beta$ with EW $>1200$~\AA\ and blue UV continua. Furthermore, although we have adopted a constant ionising photon production efficiency (i.e., hydrogen ionising photon production rate per non-ionising UV luminosity at rest-frame $1500$~\AA) for all sources regardless of luminosity, a higher value for AGN would increase their contribution to the ionising photon budget. These findings illustrate the importance of further examining lower redshift analogues such as UDS-24561. 

Given its rarity in the parent sample of extreme [O~{\small III}] emitters we matched with Chandra data, it may be challenging to find further intermediate redshift AGN similar to UDS-24561. On the other hand, it may be fruitful to focus on extreme [O~{\small III}] emitters with slightly lower EWs and redder UV slopes given the continuity in the population and the contrast with nearby AGN in Figure~\ref{fig:beta_dist}. By expanding such searches to encompass other wide-field surveys, we can hopefully provide the necessary baseline of properties of low redshift AGN for comparison with targets in the reionisation era.

Ultimately the statistical baseline of spectroscopic properties of blue, extreme [O~{\small III}] galaxies harbouring AGN at intermediate redshift including UDS-24561 will provide useful clues in guiding future {\it JWST} emission line surveys in the reionisation era. Since rest-frame UV and optical spectra of $z>6$ galaxies will be obtained by {\it JWST}/NIRSpec, AGN activities in these systems could be revealed via broad optical emission features or UV line diagnostics involving high ionisation emission lines like C~{\small IV} and He~{\small II} as discussed in Section~\ref{sec:results}. This will provide a valuable census of AGN at $z>6$, hence constraining the contribution of AGN to cosmic reionisation.


\section*{Acknowledgements}

MT and RSE acknowledge funding from the European Research Council under the European Union Horizon 2020 research and innovation programme (grant agreement No. 669253). RE acknowledges funding from JWST/NIRCam contract to the University of Arizona, NAS5-02015. We would like to thank Nicolas Laporte, Jianwei Lyu, and Feige Wang for useful discussions. 

This work is based on observations taken by the 3D-HST Treasury Program (GO 12177 and 12328) with the NASA/ESA HST, which is operated by the Association of Universities for Research in Astronomy, Inc., under NASA contract NAS5-26555. Part of the observations reported here were obtained at the MMT Observatory, a joint facility of the University of Arizona and the Smithsonian Institution. We acknowledge the MMT queue observers for assisting with MMT/Binospec observations. Part of the data presented herein were obtained at the W. M. Keck Observatory, which is operated as a scientific partnership among the California Institute of Technology, the University of California and the National Aeronautics and Space Administration. The Observatory was made possible by the generous financial support of the W. M. Keck Foundation. The scientific results reported in this article are based in part on data obtained from the Chandra Data Archive.

This research made use of {\footnotesize ASTROPY}, a community-developed core {\footnotesize PYTHON} package for Astronomy \citep{AstropyCollaboration2013}, {\footnotesize NUMPY}, {\footnotesize SCIPY} \citep{Jones2001}, and {\footnotesize MATPLOTLIB} \citep{Hunter2007}.


\section*{Data Availability}

The 3D-HST data can be accessed from \url{https://archive.stsci.edu/prepds/3d-hst/}. The data underlying this article will be shared on reasonable request to the corresponding author.



\bibliographystyle{mnras}
\bibliography{z3_AGN}



\appendix


\bsp	
\label{lastpage}
\end{document}